\documentclass[10pt]{article}
\usepackage[usenames,dvipsnames]{xcolor}
\usepackage[OE]{express_ax}

\begin{document}

\title{Engineering temporal-mode-selective frequency conversion in nonlinear optical waveguides: From theory to experiment}

\author{Dileep V. Reddy\authormark{1,*} and Michael G. Raymer\authormark{1}}

\address{\authormark{1}Oregon Center for Optical, Molecular, and Quantum Science, and Department of Physics, 1274 University of Oregon, Eugene, OR 97403, USA}

\email{\authormark{*}dileep@uoregon.edu} 



\begin{abstract}
Quantum frequency conversion (QFC) in nonlinear optical media is a
powerful tool for temporal-mode selective manipulation of
light. Recent attempts at achieving high mode selectivities and/or
fidelities have had to resort to multi-dimensional optimization
schemes to determine the system's natural Schmidt modes. Certain
combinations of relative-group velocities between the relevant
frequency bands, medium length, and temporal pulse widths have been
known to achieve good selectivities (exceeding $80$\%) for temporal
modes that are nearly identical to pump pulse shapes, even for high
conversion efficiencies. Working in this parameter regime using an
off-the-shelf, second-harmonic generation, MgO:PPLN waveguide, and
with pulses on the order of $500$ fs at wavelengths around $800$ nm,
we verify experimentally that model-predicted Schmidt modes provide
the high temporal-mode selectivity expected. This paves the way to the
implementation of a proposed two-stage QFC scheme that is predicted to
reach near-perfect ($100$\%) selectivity.
\end{abstract}


\bibliographystyle{osajnl}

\section{Introduction}

Temporal modes of light \cite{Smith:2007njp} are a burgeoning new
subspace for both classical and quantum information encoding
\cite{ptd12,Roslund:14np,Valentin:14pra,Brecht:15prx,Pant:16pra,Lenhard:17ax}. The
temporal mode (TM) bases are discrete and in principle infinite
dimensional, and truly orthogonal. These orthogonal modes have
overlapping spectra at the same carrier frequency, fully overlap in
time, and share the same polarization and transverse spatial mode
properties, making them convenient for long-distance communication and
interfacing with optical memories
\cite{Nunn:07pra,Zheng:15pra,Costanzo:16prl,Zhong:17nc}. Co-propagating
TMs maintain relative coherence across the entire mode space, and
experience the same reversible unitary dispersion during propagation
in typical media. In order to fully exploit TMs as a resource, we need
highly TM-selective, high-fidelity devices that can generate, sort,
store, measure, and generally manipulate them, even at the
single-photon level
\cite{EcksteinA:2011vg,Brecht:2011hz,Brecht:2014eg,Ansari:16ax,Manurkar:16,Manukar:16ax,Averchenko:16ax,Karpinski:17np,Ansari:17ax,Allgaier:17ax,Ra:17ax}.

Many optically-pumped processes, such as atomic-ensemble Raman
memories
\cite{Nunn:07pra,Hammerer:10rmp,Bustard:13prl,Reim:12prl,Humphreys:14prl}
and nonlinear optical frequency conversion \cite{Kumar:90} via
three/four-wave mixing \cite{Huang:92prl,mcg10a,mcg10b,ptd12}, are
already known to be TM sensitive when using pulsed pumps
\cite{Huang:13,Kowligy:14,Donohue:15pra}, and show TM discriminatory
behavior at low coupling strengths
\cite{mcg11,mej12,EcksteinA:2011vg,Reddy:2013ip}. Increasing the pump
intensities causes the TM discrimination to peak and regress to
smaller values, thus ultimately limiting the maximum selectivity of
straight-forward implementations of such processes
\cite{Reddy:2013ip}. This loss in selectivity is understood to be a
result of time-ordering effects between interacting temporal
wave-packets of the participating fields (be they electromagnetic,
polarization-phononic, acoustic, etc.) convecting through each other
at varied group velocities \cite{Reddy:2013ip,Quesada:16}. This
selectivity limit, however, can be asymtotically overcome
\cite{Reddy:2014bt} by using interferometric techniques (``TM
interferometry'') in cascaded setups to enhance TM separability up to
$100$\%, even at large coupling strengths
\cite{Reddy:2014bt,Reddy:15pra,Nunn:16ax,Quesada:16,Clemmen:16prl,Kobayashi:17ax}. This
new capability opens new avenues for information processing, including
the complete set of operations needed for linear-optical quantum
information processing \cite{Brecht:15prx}.

Quantum frequency conversion through $\chi^{(2)}$- or $\chi^{(3)}$-
nonlinear media is gaining in usefulness as an all-optical means of
implementing a fully-programmeable ``quantum pulse gate,'' whose role
is to multiplex and demultiplex TMs \cite{EcksteinA:2011vg}. The
ability of nonlinear frequency conversion to unitarily reshape, and
preserve quantum information stored in the TM basis is well known
\cite{Kumar:90,enk10,Colin:12pra}. The TM-selectivity of frequency
conversion \cite{Reddy:2013ip} suffers from the above-mentioned
time-ordering-induced upper limit, necessitating an eventual
exploration of TM interferometry. Experimental exploration of TM
discrimination is still in its nascent stages
\cite{Brecht:2014eg,Ansari:16ax,Manurkar:16,Ansari:17ax}. While the
theoretically predicted selectivity limit still stands to be broken,
other more specific predictions of said theory have yet to be
validated. In this paper, we redress this issue both experimentally
and theoretically, as well as provide a simple parameter-set guideline
to aid in designing similar processes in other systems and spectral
regions.

In the following, we first model pulsed-pump mediated,
second-order-nonlinear-optical frequency conversion between a signal
band and an idler band as a set of coupled-mode equations, and define
a TM-selectivity figure of merit to characterize the process. We then
showcase high TM selectivity in a parameter regime where the pump and
signal pulses propagate with the same group velocity, but the idler
group velocity is significantly different. We illustrate specific
predictions of the model for various input conditions, and propose a
means of adequately approximating the aforementioned parameter regime
by choosing band-carrier frequencies around the ones phase matched for
second-harmonic generation (SHG) in typical off-the-shelf nonlinear
waveguides. We then demostrate TM-selective frequency conversion in a
$5$ $\mu$m wide, $5$ mm long, MgO:PPLN waveguide periodically poled
for SHG from $816.6$ nm to $408.3$ nm at $24.25^\circ$C. We employ a
Kerr-modelocked, ultrafast titanium-sapphire laser and a folded
Treacy-grating pair pulse shaper with a reflective, 2D spatial light
modulator, to situate and shape the pump and signal bands at $821$ nm
and $812.2$ nm respectively, with bandwidths $\sim 2.5$ nm. We use
pump pulses with energies of order $10$ nJ, and pulses of temporal
widths of order $500$ fs, to verify all the predicted features of our
model. The first two numerically computed, natural Schmidt modes of
the model achieved a $4.7$-to-$1$ contrast ($85$\% vs. $18$\%) in
conversion efficiencies for the two tested pump shapes in both theory
and experiment. While the experiments are carried out using weak
coherent-state signals, theory indicates the same conversion
efficiencies and TM selectivities would be observed using heralded
single-photon wave packets \cite{mcg11}.

\section{Theory and modelling}

\subsection{Equations of motion and selectivity}

The model for frequency conversion (FC) of temporal wave-packet modes
in a $\chi^{(2)}$-nonlinear waveguide may be expressed as a pair of
coupled-mode equations involving the electric-field envelopes, treated
as quantum field operators \cite{Reddy:2013ip}. We designate the
letters $p$, $s$, and $r$ to denote electromagnetic fields within the
three participating frequency bands, namely, the pump, signal, and
idler bands, respectively. The band-central, or carrier frequencies
have to be constrained by energy conservation ($\omega_r =
\omega_s+\omega_p$), and the nonlinear-optical waveguide used for
frequency conversion is assumed to be periodically poled to ensure
proper phase matching ($\beta_r-\beta_s-\beta_p-2\pi/\Lambda=0$, where
$\beta_j$ are wavenumbers and $\Lambda$ is the poling period) for the
pulses' carrier frequencies. If the pulses in question are
sufficiently narrowband, we can ignore second- and higher-order
dispersion for all three bands, and write the coupled-mode equations
as

\begin{subequations}\label{eq1}
  \begin{align}
    (\partial_z + \beta'_r\partial_t)\hat{A}_r(z,t) = &i\gamma A_p(t-\beta'_pz)\hat{A}_s(z,t),\\
    (\partial_z + \beta'_s\partial_t)\hat{A}_s(z,t) = &i\gamma A_p^*(t-\beta'_pz)\hat{A}_r(z,t).
  \end{align}
\end{subequations}

Here, $\gamma$ is a composite parameter denoting the strength of the
interaction, and is linearly dependent on the medium nonlinearity
($\chi^{(2)}$), and the pump-pulse energy, as well as transverse
spatial mode-overlap functions. The parameters $\beta'_j \equiv
\partial_\omega\beta(\omega)|_{\omega_j}$ are the corresponding
group-slownesses (inverse group velocities) for the bands
$j\in\{r,s,p\}$, and the operators $\hat{A}_j(z,t)$ denote
field-annihilation operators for square-normalized, slowly-varying,
complex amplitudes of the temporal modes
\cite{Smith:2007njp,Brecht:15prx} within the medium at location $z$
along the propagation direction.

We require no additional noise operators in Eqs. \ref{eq1} since
quantum FC is in principle a noiseless, unitary process
\cite{Louisell:61pr}. In the single-photon case, these operators can
be replaced by quantum wavefunction amplitudes
\cite{Smith:2007njp,mcg11}. For weak coherent pulses (as in the case
of our experiment), they can be replaced by complex functions denoting
the classical pulse-envelope shapes, because the same equations
hold. The strong classical pump pulse $A_p(t)$ is assumed to be
unchanging during FC. The medium is of length $L$, and $z=0$ defines
its input face. The nonlinearity of the medium is assumed uniform
throughout its length.

For given values for all of the parameters defined above, and a known
pump-pulse shape $A_p(t)$, FC in a finite medium can be treated as a
scattering process that relates the input temporal modes
$\hat{A}_j(0,t')$ to the output modes $\hat{A}_j(L,t)$ via a set of
Green function (GF) relations:

\begin{equation}
  \hat{A}_j(L,t) = \sum\limits_{k=r,s}\int\limits_{-\infty}^{\infty}G_{jk}(t,t')\hat{A}_k(0,t')dt'.
\end{equation}

Each output field $j \in \{r,s\}$ is linearly dependent on both input
fields $k \in \{r,s\}$, where $t'$ denotes an input time and $t$
denotes an output time. This formalism is convenient for analysis of
temporal-mode selectivity \cite{Reddy:2013ip,Reddy:2014bt}, as the
separability of the four Green function subkernels can be quantified
via their singular-value decomposition \cite{enk10,stra98,gbur}:
\begin{subequations}
\begin{align}
G_{rr}(t,t') &= \sum\limits_n\tau_n\Psi_{n}(t)\psi_{n}^*(t'),\quad
&G_{rs}(t,t') &= \sum\limits_n\rho_n\Psi_{n}(t)\phi_{n}^*(t'),\\
G_{ss}(t,t') &= \sum\limits_n\tau_n^*\Phi_{n}(t)\phi_{n}^*(t'),\quad
&G_{sr}(t,t') &= -\sum\limits_n\rho_n^*\Phi_{n}(t)\psi_{n}^*(t').
\end{align}
\end{subequations}

The functions $\psi_{n}(t'), \phi_{n}(t')$ are the input ``Schmidt
modes'' and $\Psi_{n}(t), \Phi_{n}(t)$ are the corresponding output
Schmidt modes for the {\it r} and {\it s} bands respectively. The
Schmidt modes define the sets of ``natural temporal modes'' for the FC
problem given a particular pump shape and medium characteristics. For
a given integer index $n$, the quartet of modes are related to each
other in a beamsplitter-like transformation through the Schmidt
coefficients ($\tau_n$, $\rho_n$), which obey the unitarity
constraint, $|\tau_n|^2+|\rho_n|^2=1$. In simple terms, if the input
state in the $s$-band were to be a single photon (or a weak, coherent
pulse) in the temporal mode $\phi_n(t')$, and the $r$-band input were
to be vacuum, then the probability (efficiency) of frequency
conversion from $s$-band to $r$-band would be $|\rho_n|^2$. The
converted component exits in the $r$-band in mode $\Psi_n(t)$, and the
unconverted component (occurring with probability/efficiency
$|\tau_n|^2$) exits in the $s$-band in the mode $\Phi_n(t)$. Note that
each of the four sets of Schmidt modes forms a complete basis set for
its band and input/output context. The effect of FC on an arbitrary
input temporal mode can be easily computed by expressing said input
state in the natural input Schmidt-mode basis of the device and
employing the Schmidt-coefficient beamsplitter relations.

We choose to arrange the Schmidt modes in the four sets, and the
corresponding Schmidt coefficients, in decreasing order of conversion
efficiency (CE), such that $|\rho_1|^2 \ge |\rho_2|^2 \ge |\rho_3|^2
...$ and so on. An FC device with perfect mode discrimination would
have a non-zero $\rho_1$, and $\rho_{n>1}=0$, meaning the GF subkernel
is separable but not necessarily 100\% efficient. A device with
perfect mode {\it selectivity} would have $\rho_j=\delta_{j,1}$. To
characterize mode-selectivity, we define a figure of merit called
selectivity $S = |\rho_1|^4/\sum_{n=1}^\infty |\rho_n|^2$
\cite{Reddy:2013ip}. It has previously been shown that perfect
selectivity cannot be achieved in simple inter-pulse interaction
systems due to time-ordering corrections \cite{Quesada:16}. But
selectivity asymptotically approaching $100$\% can be achieved in
cascaded, multi-stage FC implementations
\cite{Reddy:2014bt,Reddy:15pra,Christensen:15,Quesada:16}. This paper
does not address multi-stage FC.

\subsection{Group-velocity matched regime}

We can condense all of the model parameters into three dimensionless
quantities in order to aid mapping settings and results from diverse
FC systems to this model. These are \cite{Reddy:2013ip}:

\begin{equation}
  \widetilde\gamma = \gamma\sqrt{\frac{L}{\beta'_{rs}}},\quad \zeta=\frac{\beta'_{rs}L}{\tau_p},\quad \text{and }\quad\xi=\frac{\beta'_{pr}}{\beta'_{ps}},\label{eq4}
\end{equation}
\noindent where $\tau_p$ is the temporal-width of the pump pulse, and
$\beta'_{jk} = \beta'_j-\beta'_k$. $\widetilde\gamma$ is an interband
coupling strength. $\zeta$ is the signal-idler inter-pulse walk-off
relative to pump width, and $\xi$ is the group-velocity mismatch
contrast.

Through an exhaustive numerical exploration \cite{Reddy:2013ip}, we
have previously determined that for good GF separability at low pump
energies, as well as the best selectivity ($\sim 0.83$) at higher pump
energies, the best parameter regime is $\xi \gg \zeta \gg 1$. By
designing the system such that the group-velocity of the pump pulse is
identical to that of one of the other bands (the $s$-band, for the
definitions in Eq. \ref{eq4}), and highly different from that of the
remaining ($r$) band, we can have $\xi\rightarrow\infty$. We call this
condition the group-velocity matched (GVM) regime, and it was first
considered in detail in \cite{EcksteinA:2011vg}. Very large $\zeta$,
however, are physically inaccessible since realistic nonlinear
waveguides have finite length, and the pump pulse needs to remain
reasonably narrowband to maintain phase-matching and avoid
higher-order dispersion.

The Green function solutions for the GVM regime are known in closed
analytical form \cite{Colin:12pra,Reddy:2013ip}, which aids in
numerical analysis, as well as physical system design. We have
validated these analytical solutions using regime-agnostic wave-mixing
simulations based on a numerical split-step implementation of the
propagation and interaction of the various fields
\cite{Reddy:2013ip,Reddy:2014bt,Reddy:15pra,Christensen:15}. Specifically,
we propagate the fields and apply dispersion in the Fourier domain,
and ``mix'' them using fourth-order Runge-Kutta in the time domain,
alternating between the two for every iteration. We compute the Green
functions by running the simulation for a basis set of input
conditions, and computing the overlap of the resulting outputs with
another spanning basis set of functions. The numerical simulations
verify that small deviations from the assumed conditions (GVM regime,
absense of higher-order dispersion, etc.) do not cause significant
departures from the predictions of the analytical solutions. We
present some of these predictions here.

\begin{figure*}[htb]
\centering
\includegraphics[width=\linewidth]{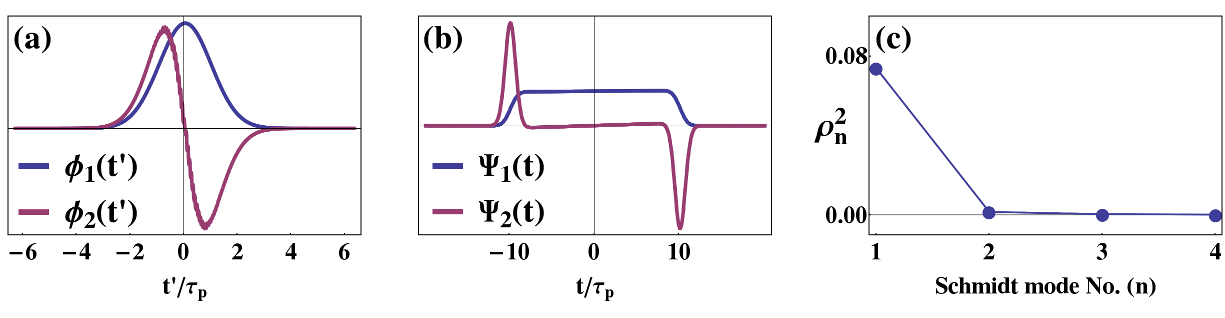}
\caption{Numerically simulated results for $\widetilde\gamma = 0.141$, $\zeta = 20$, and $\xi = \infty$ (perfect group-velocity matching) with Gaussian-shaped pump pulse. (a) The first two $s$-band input Schmidt modes. (b) The first two $r$-band output Schmidt modes. (c) Conversion efficiencies of first four dominant Schmidt modes. Note that $\phi_1(t')$ closely resembles the pump shape, and the temporal width of $\Psi_1(t)$ is larger by a factor of $\zeta = 20$.}
\label{fig01}
\end{figure*}

\begin{figure*}
\centering
\includegraphics[width=\linewidth]{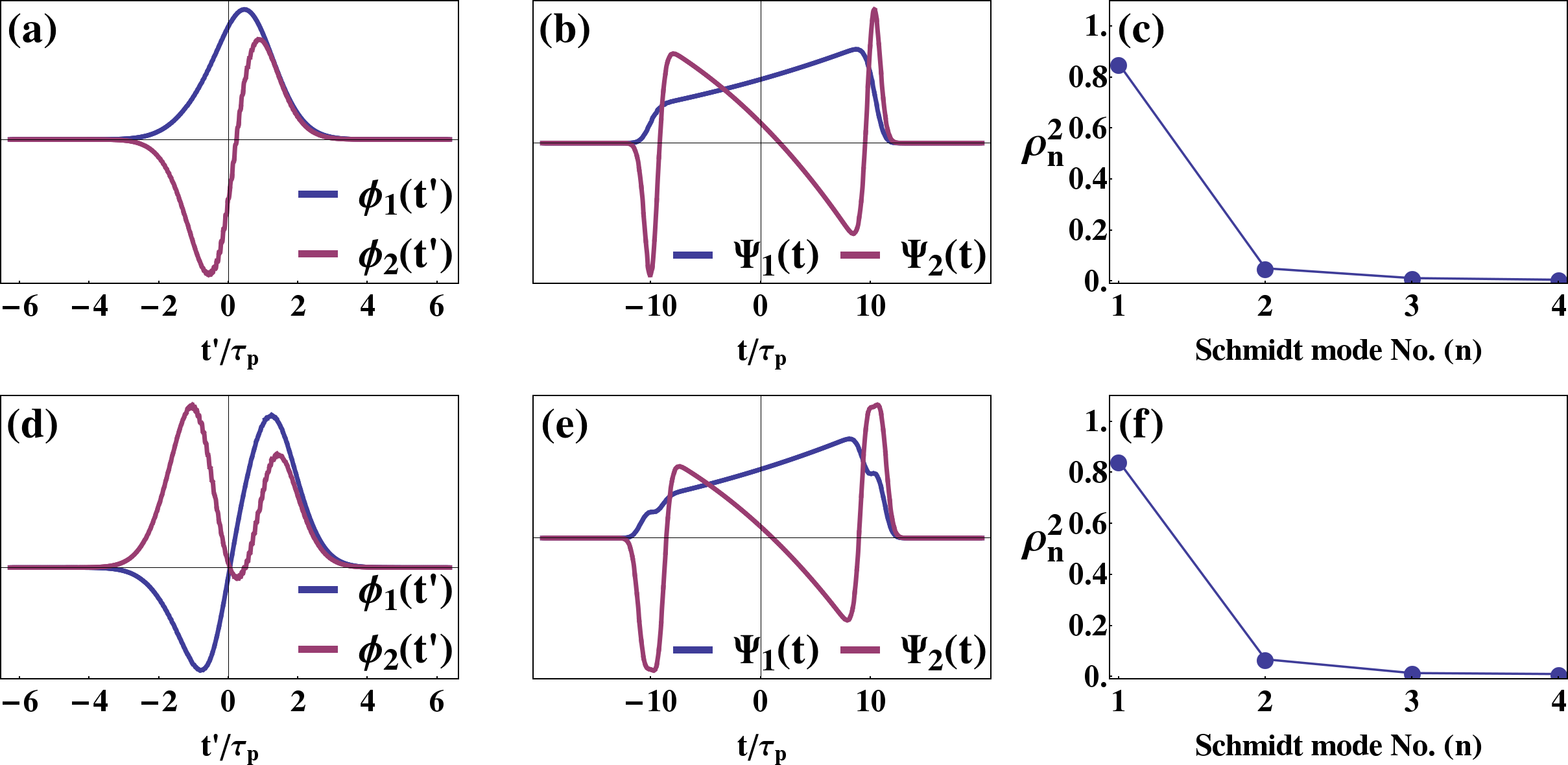}
\caption{Numerically simulated, dominant Schmidt modes and conversion efficiencies for $\widetilde\gamma = 0.707$, $\zeta = 20$, and $\xi = \infty$ (perfect group-velocity matching) with Gaussian (a, b, c) and first-order Hermite-Gaussian (d, e, f) pump pulses. Due to the lack of any complex phase structure in the pumps, the Schmidt modes end up being real valued (up to overall phase). Note that the first ($n=1$) $s$-band input Schmidt modes (a, d) resemble the  group-velocity matched, pump-pulse shapes up to temporal skewing, whereas the first $r$-band output Schmidt modes (b, e) get stretched relative to pump width by factor $\zeta$. Also note the independence of the dominant Schmidt coefficients (c, f) from pump-pulse shape.}
\label{fig02}
\end{figure*}

In the GVM regime, the group-velocity matched signal ($s$-band)
copropagates with the pump pulse in the medium, and the unmatched
idler ($r$-band) pulse falls behind (or, in the case of anamolous
dispersion, speeds ahead) of the other two pulses. Consequently, as
long as $\zeta \gg 1$ in addition to the GVM condition, both the input
and output $s$-band Schmidt modes ($\phi_j(t')$, $\Phi_j(t)$) will
have the same temporal widths as that of the pump pulse, and the
$r$-band input and output modes ($\psi_j(t')$, $\Psi_j(t)$) get
temporally stretched by a factor of $\zeta$.

In Fig. \ref{fig01}, we plot the first two $s$-band input Schmidt
modes and the first two $r$-band output Schmidt modes for a low
coupling strength ($\widetilde\gamma = 0.141$). The temporal
stretching effect between the mode widths of the two bands by a factor
of $\zeta = 20$ is clearly illustrated. Also note in
Fig. \ref{fig01}(c) that although the CE of the first Schmidt mode is
miniscule, it is large compared to that of the second Schmidt
mode. The target TM (for optimum FC), which is the first $s$-input
Schmidt mode in Fig. \ref{fig01}(a), is nearly identical to the pump
shape, while the second mode is temporally orthogonal to the first.

Figure \ref{fig02} shows the same data as in Fig. \ref{fig01}, but for
a higher coupling strength ($\widetilde\gamma = 0.707$), and two
different pump-pulse shapes: Gaussian (Fig. \ref{fig02}(a,b,c)), and
first-order Hermite Gaussian (Fig. \ref{fig02}(d,e,f)). Note that the
CE of the second Schmidt mode is no longer negligible.

\begin{figure*}[htb]
\centering
\includegraphics[width=\linewidth]{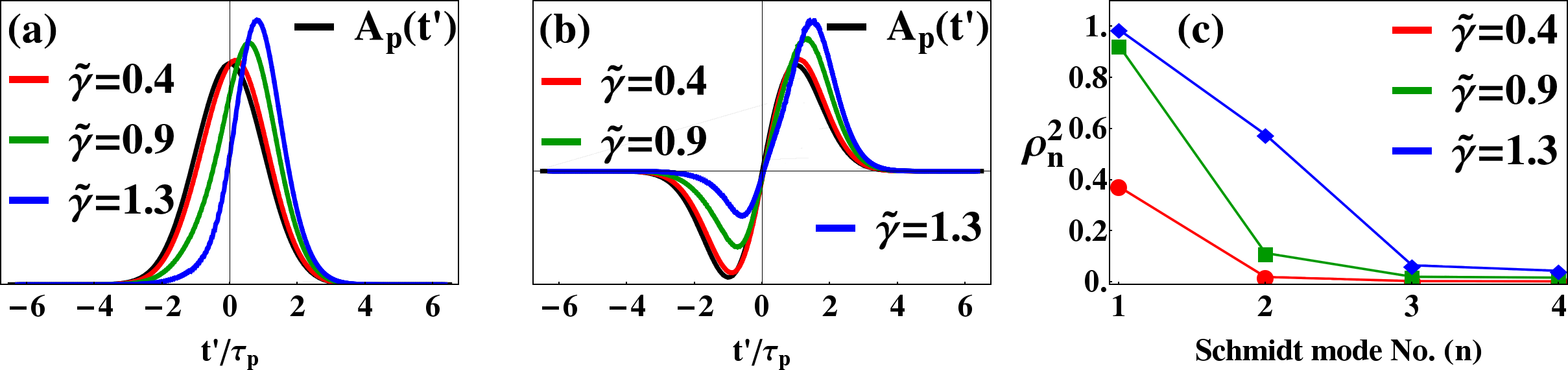}
\caption{Numerically simulated, first ($n=1$) $s$-band input Schmidt modes ($\phi_1(t')$) for $\zeta = 20$, $\xi = \infty$, and various $\widetilde\gamma$ with (a) Gaussian pump pulses, and (b) first-order Hermite-Gaussian pump pulses. These plots demonstrate the $\widetilde\gamma$-dependent temporal skewing effect. (c) The conversion efficiencies for the first four dominant Schmidt modes. Plot (c) is identical for both pump-pulse shapes.}
\label{fig03}
\end{figure*}

In the GVM regime, even at large pump-pulse energies (coupling
strengths), the complex shape of the pump-pulse envelope fully
determines the shape of the dominant $s$-band Schmidt modes
(Fig. \ref{fig02}(a,d)). The dominant $r$-band Schmidt modes, however,
are influenced very little by the pump shape, and instead reflect the
variation of the medium nonlinearity along the waveguide propagation
direction (Fig. \ref{fig02}(b,e)), which in our case is considered to
be uniform. At low pump-pulse energies, or small $\widetilde\gamma$,
the first ($n=1$) $s$-band input and output Schmidt modes are nearly
identical to the pump pulse shape (and the first $r$-band modes look
like flat, square pulses). But at higher $\widetilde\gamma$, the first
Schmidt mode changes into a temporally skewed version of the pump
pulse (see Fig. \ref{fig03}). The Schmidt coefficients (and therefore,
the Schmidt-mode CE) are independent of the pump-pulse shape. Figure
\ref{fig03}(c) depicts the rapid rise of the CEs of the second and
higher Schmidt modes with increasing $\widetilde\gamma$, which results
in a cap on the maximum achievable selectivity.

Curiously, for a given $\widetilde\gamma$, different orthogonal
pump-pulse shapes would result in mutually orthogonal dominant
$s$-band Schmidt modes, even if the modes cease to resemble the
original pump shapes.

In our experiment, we validate these precise predictions by following
two approaches. Firstly, for given pump-pulse shape and energy, we
maximize the signal CE by attempting to match the $s$-input Schmidt
modes predicted by theory. We numerically precompute the first
Schmidt-mode shape for both Gaussian and Hermite-Gaussian pump pulses
for a continuous range of $\widetilde\gamma$. Then the problem of
matching the Schmidt mode at a given pump power reduces to CE
maximization through a scan of the single parameter
$\widetilde\gamma$.

\begin{figure*}[htb]
\centering
\includegraphics[width=\linewidth]{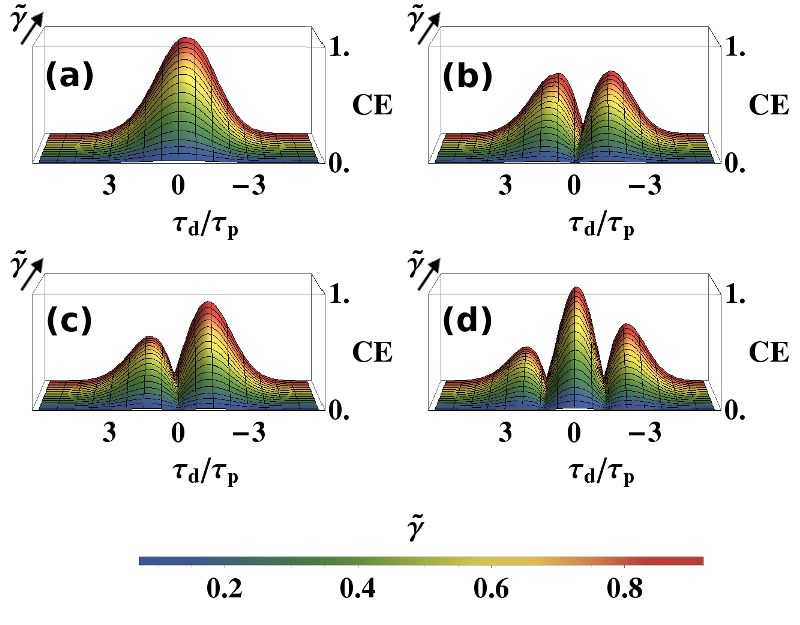}
\caption{Numerically simulated, 3D surface plots of conversion efficiencies (CE) vs. input pump-signal time delay by pump width ($\tau_d/\tau_p$) for various $\widetilde\gamma$ ($\propto\sqrt{P_{\text{pump}}}$) for (a) Gaussian pump and signal, (b) Gaussian pump and first-order Hermite-Gaussian signal, (c) first-order Hermite-Gaussian pump and Gaussian signal, and (d) first-order Hermite-Gaussian pump and signal pulse shapes. Note the temporal skewness at higher $\widetilde\gamma$, reflected in shift of CE maxima with respect to mesh grid, as well as asymmetry in lobe peak heights.}
\label{fig04}
\end{figure*}

A second means of verifying the model is to keep the signal input
pulse shapes static, but delay them with respect to the pump pulse and
chart the CE. For Gaussian and first-order Hermite-Gaussian pump and
signal shapes, four surface plots of CE for various input
inter-pump-signal delays and pump energies have been numerically
generated and plotted in Fig. \ref{fig04}. The $s$-band Schmidt mode
distortions show up as temporal shifts and lobe-peak asymmetries at
higher $\widetilde\gamma$.

\subsection{Physical system design and parameter selection}

In order to take advantage of theoretical predictions, one must choose
the physical system parameters to match the regimes of applicability
of said theory. In the context of TM-selective frequency conversion,
the GVM or near-GVM condition ($\xi \gg \zeta \gg 1$) must constrain
the selection of waveguide material, length, band central frequencies,
and TM bandwidths. At the time of this writing, only two other groups
\cite{Brecht:2014eg,Ansari:16ax,Manurkar:16,Ansari:17ax} have
addressed experimental frequency conversion in a TM-selective context.

The Silberhorn group \cite{Brecht:2014eg,Ansari:16ax,Ansari:17ax} use
homebuilt, periodically-poled lithium niobate (PPLN) waveguides with a
poling period of about $4.4$ $\mu$m. Such a period gives the optimum
phase-matching for sum frequency generation (SFG) from bands centered
near $1550$ nm and $860$ nm, into the band around $550$ nm. They
engineer their waveguide dispersion to achieve perfect GVM
($\xi\rightarrow\infty$) at these wavelengths, and compensate for
fabrication errors by tuning the waveguide temperature in the
$150-200^\circ$C range. They can thus afford to use longer waveguides
($\sim 17-27$ mm) and short pulse lengths ($\sim 200$ fs) and obtain
large $\zeta$ values without having to worry about signal-pump
inter-pulse walkoff within the medium during propagation.

Although exact GVM is optimum for TM-selectivity, one can deviate from
perfect signal-pump GVM and still retain most of its advantages, as
long as $\xi \gg \zeta \gg 1$ is satistied. The Kumar/Kanter group
\cite{Manurkar:16} hit upon an interesting solution that allows for
minor deviation from GVM. They used a $52$ mm long, custom PPLN
waveguide designed for second-harmonic generation (SHG) from $1544$ nm
into $772$ nm at $73.4^\circ$C, and situated their pump ($1556.6$ nm)
and signal ($1532.1$ nm) bands symmetrically on either side of the SHG
wavelength, yielding $\xi\approx 215$. Their pump/signal sources and
pulse shapers restricted their temporal widths to around $5$ ps,
implying $\zeta\approx 3$. The theory predicts
\cite{Reddy:2013ip,Christensen:15} that their selectivity would
improve significantly with larger pump/signal bandwidths, as we have
used here.

\begin{figure*}[htb]
\centering
\includegraphics[width=\linewidth]{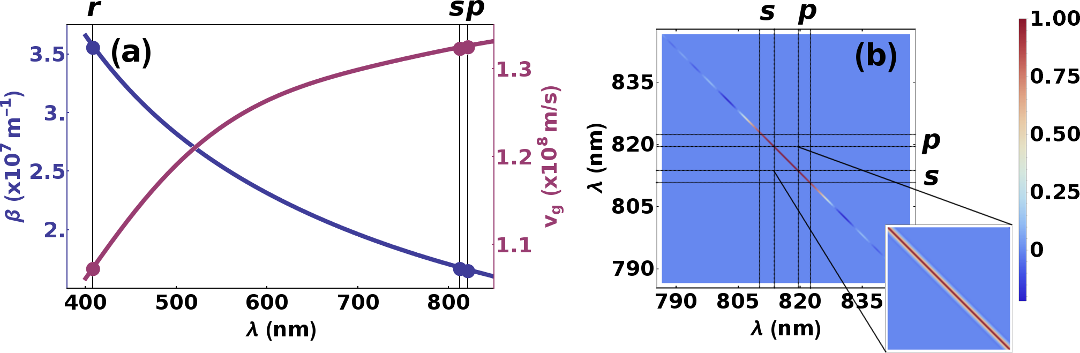}
\caption{(a) The wavenumber ($\beta$) and the group velocity ($v_g = d\omega/d\beta$) vs. wavelength ($\lambda$) for a typical $5$ $\mu$m wide, periodically-poled, MgO:LN waveguide. Also shown are the $r$-, $s$-, and $p-$ bands that we utilize for SFG. (b) Numerically computed, peak normalized joint-spectral amplitude of the degenerate, Type-0 SPDC photon pairs that would be generated in $5$ mm of such a waveguide when pumped with $0.1$ nm wide blue light in the $r$-band. Also shown are the signal ($s$) and pump ($p$) bands for the SFG process, which are situated symmetrically on either side of the red second-harmonic generation pump wavelength at $816.6$ nm. Due to the frequency anti-correlatedness of the SPDC joint-spectral amplitude, both $s$- and $p$-bands need to contain non-zero optical energies for SFG to occur into the $r$-band.}
\label{fig05}
\end{figure*}

We also employ an SHG waveguide ($816.6$ nm to $408.3$ nm at
$24.25^\circ$C) for FC by situating the signal and pump bands on
either side of the red SHG pump band. Typical SHG acceptance
bandwidths are very narrow. As long as the FC-pump band is
sufficiently detuned from the SHG-pump wavelength, so as to avoid
pump-only spurious blue-light generation, the relative spectral
flatness of normal dispersion ensures near-GVM conditions (see
Fig. \ref{fig05}(a)). One must choose a temporal width that is small
enough to ensure $\zeta\gg 1$ (for large idler-pump walkoff), but wide
enough to ensure that the pump-signal inter-pulse walkoff within the
medium remains a small fraction of the total pulse widths
($\xi\gg\zeta$). SHG waveguides, when pumped at the sum frequency can
generate degenerate photon-pairs via spontaneous parametric down
conversion (SPDC). The joint-spectral amplitude of the pairs are
tightly anticorrelated in frequency (Fig. \ref{fig05}(b)), reflecting
the narrowness of the SHG red-pump acceptance band
\cite{Laiho:09}. But the individual photons of the pair would be
wideband, allowing for sum frequency generation from two highly
detuned frequency bands on either side of the SHG-pump
wavelength. Here, SFG is really a band-restricted inverse of
SPDC. This behavior makes off-the-shelf waveguides suitable for
TM-selective FC experiments.

\section{Experiment}

\subsection{Apparatus}

\begin{figure*}[htb]
\centering
\includegraphics[width=\linewidth]{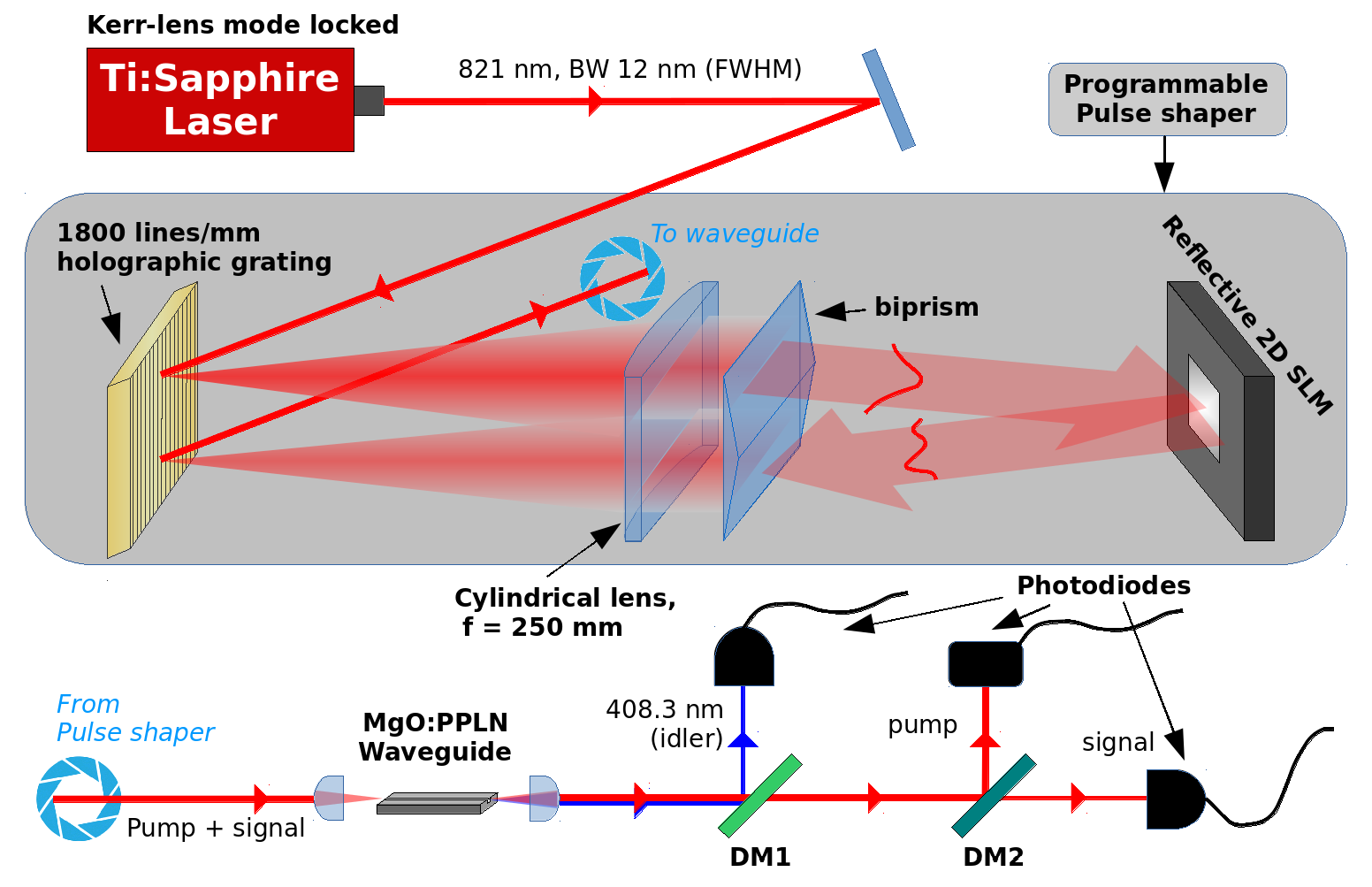}
\caption{Experimental setup. The holographic grating was used in near-Littrow mode for both incoming and outgoing beams. The $m=1$ order reflection from two separated vertical blazed gratings rendered on the SLM was recombined on the holographic grating. Both the input and output couplers of the $5$ $\mu$m wide, $5$ mm long MgO:PPLN waveguide were single-element aspheric lenses of focal length $f = 11$ mm. DM stands for dichroic mirror. Some frequency filters are not shown.}
\label{fig06}
\end{figure*}

We derived both our strong pump and weak signal pulses by reshaping
the output of a homebuilt, Kerr-lens mode-locked, ultrafast Ti:Sapph
laser with pulse duration of about $80$ fs. We tuned the cavity to
cause it to lase at $821$ nm with a FWHM bandwidth of around $12$ nm,
at a pulse rate of $76$ MHz. This beam was spatially expanded to a
transverse width of $\sim 10$ mm and sent into a folded,
4$f$-configured Treacy-grating-pair \cite{Weiner:00} pulse shaper,
which uses a reflective spatial-light modulator (SLM) in its Fourier
plane (Fig. \ref{fig06}). The pulse shaper utilized a $1800$ lines/mm
holographic grating in near-Littrow mode, and a cylindrical lens of
focal length $250$ mm to focus the wavelets onto the SLM. The lens is
cylindrical in order to spread the beam intensity vertically so as to
avoid damaging the SLM. This gave us a horizontal spot size of $\sim
30$ $\mu$m for a given wavelength. We used a custom-made biprism to
change the height of the forward and reflected light to keep the paths
symmetric, whilst sacrificing exact normal incidence on the SLM
pixels, which are designed for normal incidence.

\begin{figure*}[htb]
\centering
\includegraphics[width=\linewidth]{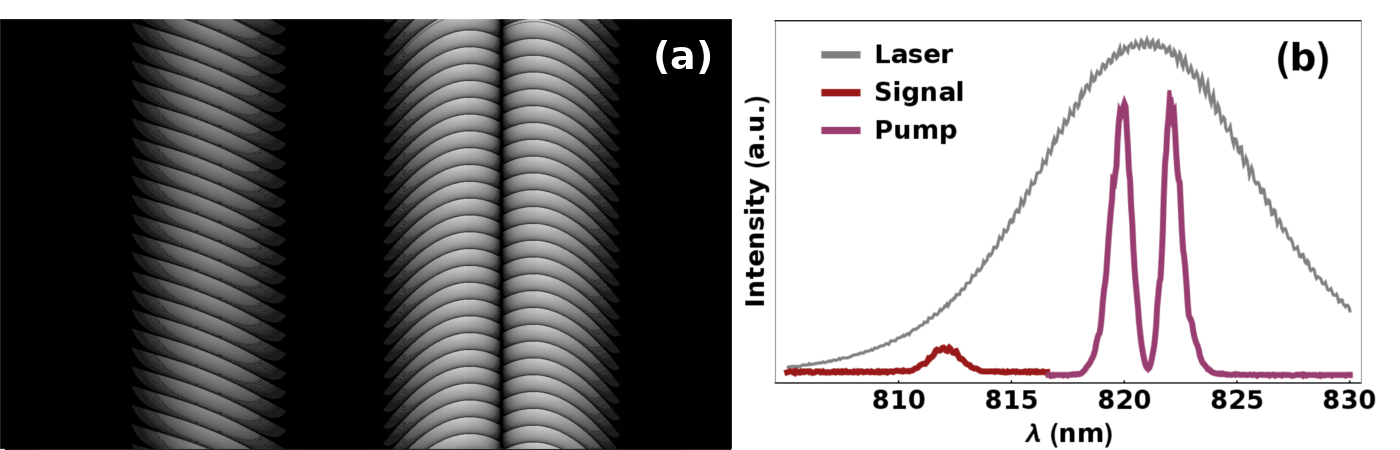}
\caption{(a) Typical 8-bit grayscale phase mask applied to the SLM to generate a Gaussian shaped signal pulse (left band) and a first-order Hermite-Gaussian pump pulse (right band). Here, the horizontal coordinate maps to different wavelengths. The phase-contrast of the vertical gratings determines amplitude. Vertically shifting the gratings can affect phase (note relative shift between the two grating patterns generating the two frequency lobes of the pump). The curved pattern is for chirp compensation (measured using a commercial FROG/GRENOUILLE 8-50-USB), and the linear spectral phase on the signal shifts it in time relative to the pump. (b) Spectra of original Ti:Sapph laser, the signal, and the pump (first-order Hermite Gaussian, for example) generated by the SLM phase mask in (a). The three different spectra were captured under different conditions and hence, the relative heights are not to scale.}
\label{fig07}
\end{figure*}

For the SLM, we used a Meadowlark 8-bit, 2D, phase-only liquid-crystal
spatial light modulator of $1920\times 1152$ pixel resolution and array size
of $17.6$ mm $\times 10.7$ mm. The pixels were squares of size $9.2$
$\mu$m and the fill factor was 95.7\%. The spatial dispersion of the
shaper at the SLM was $0.011$ nm/pixel, although, the actual shaper
resolution is limited by the spot size. In order to modulate both
amplitude and phase, we used Silberberg group's \cite{Frumker:07}
first-order approach, where we form a vertical blazed grating pattern
on the SLM and pick off its $m=1$ reflection as the output. Different
phase ramps may be applied to different wavelengths (at different
horizontal positions) to affect the amount of power in the $m=1$
reflection, and the phases can be manipulated by vertically shifting
the blazed grating upwards/downwards. We used a vertical period of
$44$ pixels in the pump band, and $50$ pixels in the signal band, as
shown in Fig. \ref{fig07}.

For the bandwidths we chose to work with ($2-2.5$ nm), the pump band
had sufficient power to allow for a frequency-resolved-optical-grating
(FROG) measurement. A commercial GRENOUILLE 8-50-USB by Swamp Optics
was employed for this. This allowed us to characterize and compensate
for the frequency chirp suffered through traversal of the beam through
multiple optical elements by applying quadratic and quartic phase
corrections on the SLM phase mask (Fig. \ref{fig07}(a)). The resulting
reduction in the pulse's temporal width was independently validated on
a homebuilt autocorrelator. Similar chirp compensation was optimized
for the signal band by maximizing the CE at a low pump power in the
nonlinear waveguide. Some linear spectral phases were added to one or
both bands to overlap the pulses in time. The bandwidths for both
bands were small enough to overcome any adverse effects from
transverse-spatial chirp or pulse-front tilt induced in the beam by
the shaper.

\begin{figure}[htb]
\centering
\includegraphics[width=0.6\linewidth]{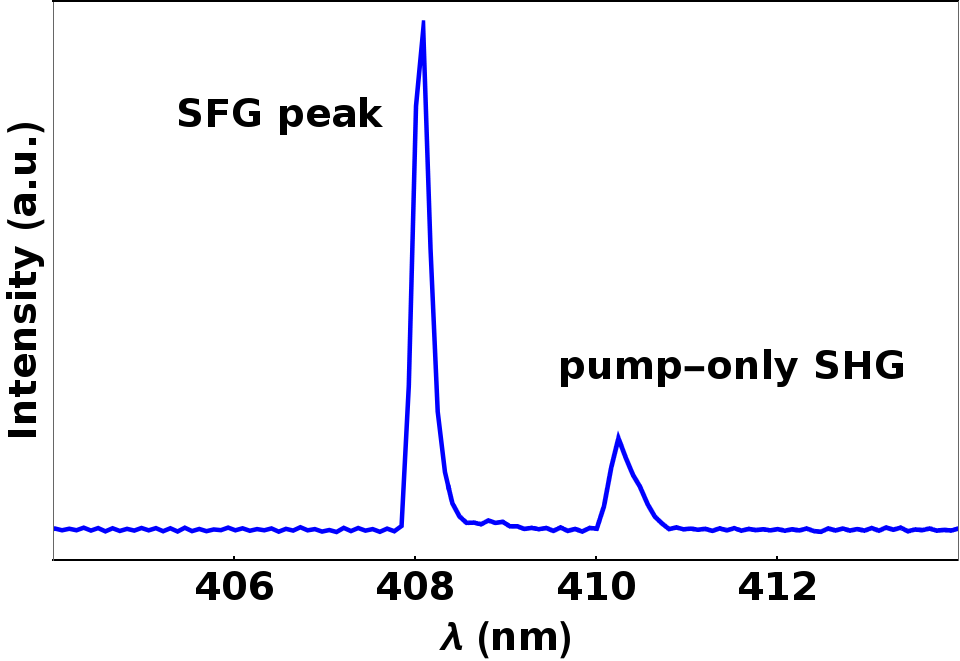}
\caption{Blue light spectra generated from the waveguide in a typical run. The SFG peak requires both the pump and signal to be present at the input, whereas the pump-only SHG peak remains even without the signal, and occurs due to imperfections in poling. The latter peak can compete with the former at higher pump powers. To use very weak signals (say sub-single-photon level), very tight spectral filtering will be needed at the blue output arm.}
\label{fig08}
\end{figure}

For frequency conversion, we used a standard, off-the-shelf, $5$ mm
long, SHG chip made of MgO-doped lithium niobate from AdvR. It had
been poled at a period of $\sim 3.1$ $\mu$m for Type-0 SHG from
wavelengths around $810-820$ nm, and consisted (as is typical) of
several waveguides of varying widths. We chose to work with the $5$
$\mu$m waveguide, which at a temperature of $24.25^\circ$C
phase-matched SHG from $816.6$ nm to $408.3$ nm. The waveguide was
placed in a homebuilt oven whose temperature was controlled using a
PID circuit. We confirmed the narrowness of the SHG red-pump
acceptance bandwidth ($< 0.5$ nm) as well as the wide SFG acceptance
bandwidth ($> 15$ nm, see Fig. \ref{fig05}(b)) by scanning the
frequencies of single-band and dual-band red inputs generated from the
pulse shaper \cite{Laiho:09} to generate blue light.

We coupled light into and out of the waveguide using $f=11$ mm
aspheric lenses, which after some post-pulse-shaper beam resizing gave
us a red-light coupling efficiency of about $30$\%. The blue and red
beams were separated at the output by a Thorlabs DMLP650 longpass
dichroic mirror, and the pump and signal bands in the red beam were
split by angle-tuning two Semrock FF01-810/10 bandpass
filters. Although the SHG/SFG process for the phase-matched
frequencies dominates when controlled for input powers, imperfections
in the waveguide resulted in spurious SHG blue light at all ``red''
wavelengths. Our process used a strong pump, but a weak signal, which
generated some pump-induced spurious SHG (Fig. \ref{fig08}) at higher
pump powers, thus possibly violating the unchanging, undepleting pump
approximation. The SFG and spurious-SHG blue bands were separated by
an angle-tuned Semrock TBP01-400/16 bandpass filter.

\subsection{Measurements}

The choice of central wavelengths for the pump ($821$ nm) and the
signal ($812.2$ nm) bands afforded us $\xi > 200$. The pump-pulse
width was set by the pulse shaper to be $\sim 530$ fs, yielding $\zeta
\approx 20$, landing us well within the near-GVM regime. The pulse
shaper allowed us a sufficient range for time shifting the pump and
signal pulses independently of each other. The average signal powers
were chosen around $20-40$ $\mu$W (measured at waveguide output),
which for a laser pulse rate of $76$ MHz, translates to $0.26-0.53$ pJ
per pulse. The pump power coupled into the waveguide was varied from
$0$ to $\sim 3.5$ mW ($46$ nJ per pulse), which was sufficient for
significant CE \cite{Christ:2013fg} without much pump depletion via
spurious SHG. In order to compare theoretical predictions with
experimental data, we needed to map the square-root of the pump power
to $\widetilde\gamma$ through a proportionality factor $\sigma$. We
fit all the diverse data for different input pulse-shape combinations
and inter-pulse delays to a single $\sigma$ value of $\sim 18$
$/\sqrt{\text{W}}$.

\begin{figure}[htb]
\centering
\includegraphics[width=\linewidth]{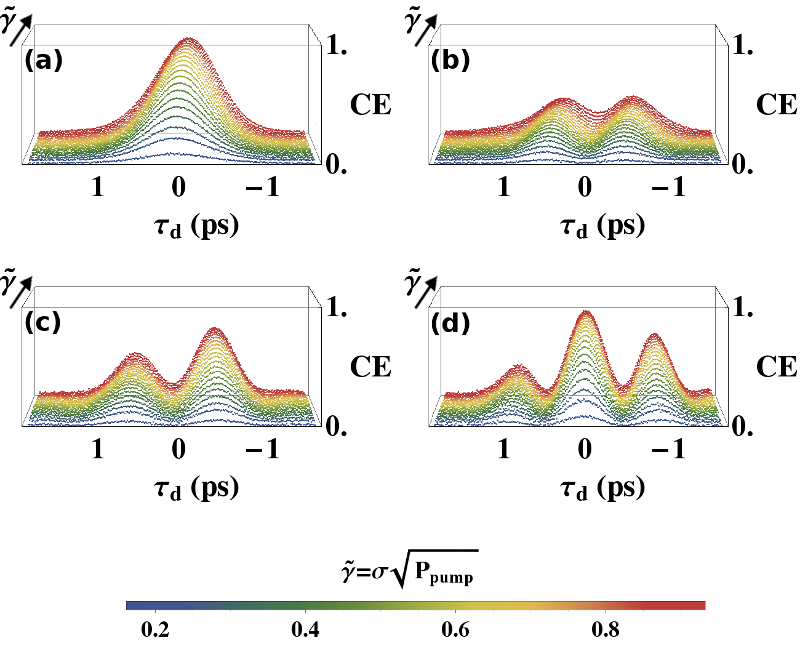}
\caption{Experimental data: 3D point plots of conversion efficiencies (CE) vs. input pump-signal time delay ($\tau_d$) for various $\widetilde\gamma = \sigma\sqrt{P_{\text{pump}}}$ (where $\sigma = 18$ /$\sqrt{W}$, and $P_{\text{pump}}$ is average pump power) for (a) Gaussian pump and signal, (b) Gaussian pump and first-order Hermite-Gaussian signal, (c) first-order Hermite-Gaussian pump and Gaussian signal, and (d) first-order Hermite-Gaussian pump and signal pulse shapes. Note the temporal skewness at higher $\widetilde\gamma$, as well as asymmetry in lobe peak heights, matching the theoretically predicted trends from Fig. \ref{fig04}. Vertical error bars are all of order $10^{-3}$, not shown.}
\label{fig09}
\end{figure}

Before attempting to create ideal input signal shapes matched to the
system's Schmidt modes, we first studied the system's behavior when
both the pump and signal inputs were Gaussian and/or first-order
Hermite Gaussian. Figure \ref{fig09} shows the conversion efficiencies
recorded for Gaussian- and (first-order) Hermite-Gaussian-shaped pump
and signal input pulses for various pump powers and initial
pump-signal time delays ($\tau_d$). The pump powers were changed by
changing the phase-contrast of the vertical gratings in the pump band
on the SLM. The pump-signal time delay was scanned in steps of $\sim
18.3$ fs by applying a linear spectral phase ramp to the signal band
on the SLM and changing its slope. The data reproduces the broad
features predicted by theory in Fig. \ref{fig04}, namely, the
temporal-shift (both extent and direction) of the peaks and troughs
for the various shape combinations, as well as the numbers and
relative heights of the peaks. This scan ensures that we aren't seeing
an artificial contrast in CE between pump-signal shape-matched
vs. shape-mismatched cases owing to a setting dependent, systemically
applied, extreme time delay between the pump and signal input
pulses. The vertical error bars are all of order $10^{-3}$, and are
not shown for the sake of clarity.

\begin{figure}[htb]
\centering
\includegraphics[width=0.6\linewidth]{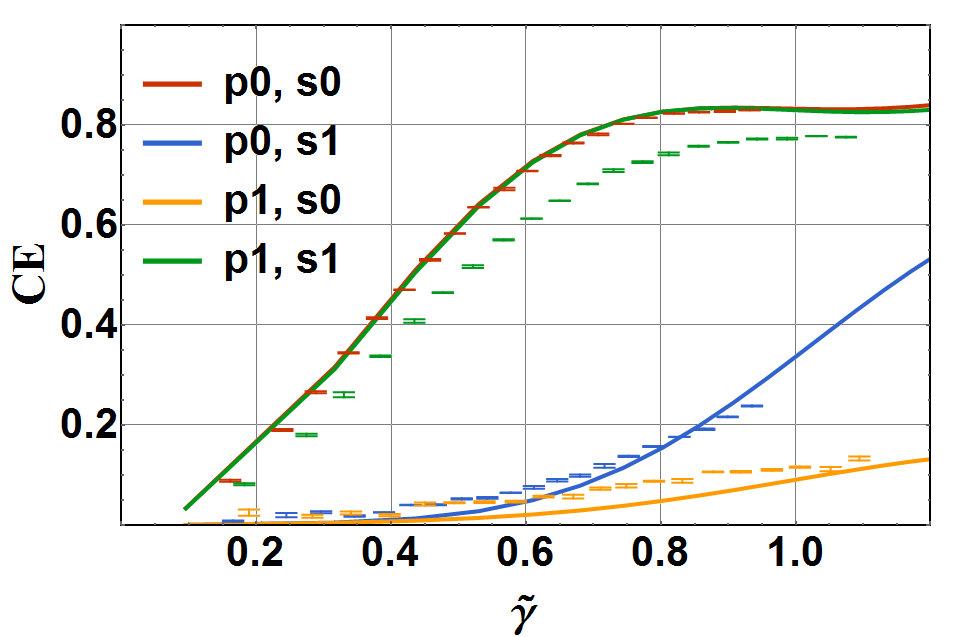}
\caption{CE (conversion efficiency) vs. $\widetilde\gamma$
  ($\propto\sqrt{P_{\text{pump}}}$) for various pump and signal input
  pulse shapes at a fixed ``zero'' delay (defined as delay that
  maximizes CE at low pump power). The legend label (p$j$, s$k$)
  denotes $j$-th order Hermite-Gaussian pump pulse, and $k$-th order
  Hermite-Gaussian signal pulse. The solid lines are theory and the
  error bars are measured data.}
\label{fig10}
\end{figure}

For closer comparison, we take a $\tau_d = 0$ slice of the theoretical
graphs and the measured data from Figs. \ref{fig04} and \ref{fig09}
respectively, and plot them in Fig. \ref{fig10}. Note that the four
possible input shape configurations follow the expected contrasts in
CE. The data points for the first-order Hermite-Gaussian-shaped pumps
are shifted horizontally forward relative to those for the
Gaussian-shaped pumps. This is because for a given temporal-width
scale, the first-order Hermite Gaussian spectrum has a slightly larger
bandwidth, giving us more available power to be syphoned off from the
ultrafast seed laser. Also note that for a given pump shape, the
signal-shape matched points are shifted horizontally slightly backward
relative to the signal-shape mismatched points, and the shift is
larger at higher CE. This is because, due to energy conservation, some
amount of power from the pump pulse is lost along with depletion of
signal power during FC. The effect, a violation of the undepleting
pump approximation, is negligible for weak signals, as demonstrated by
the close match of the data with theory.

\begin{figure}[htb]
\centering
\includegraphics[width=0.6\linewidth]{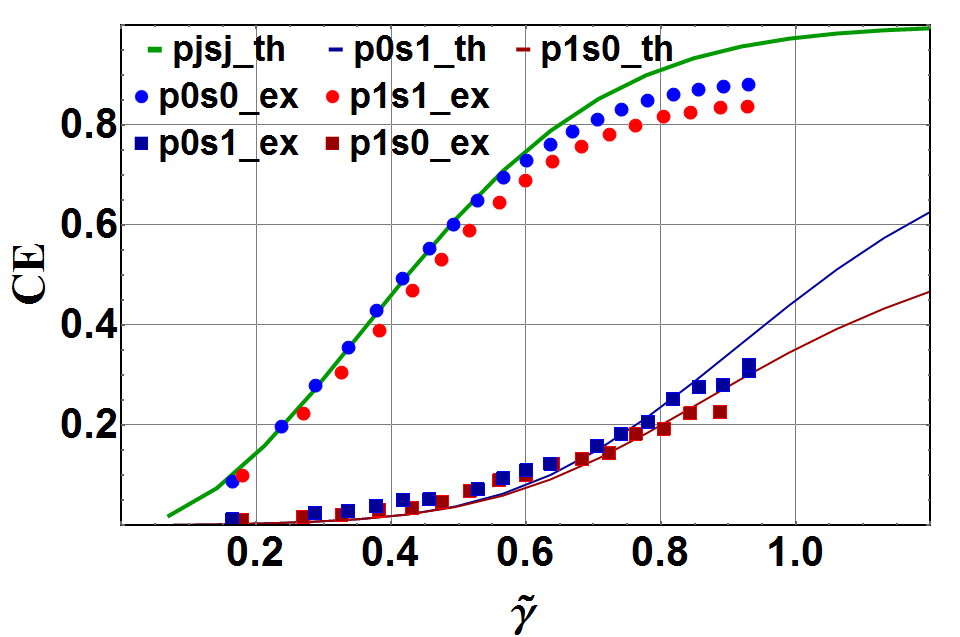}
\caption{CE (conversion efficiency) vs. $\widetilde\gamma$ ($\propto\sqrt{P_{\text{pump}}}$) for $j$-th order Hermite-Gaussian pump pulses ($pj$) and the corresponding first input Schmidt modes ($sj$). Also shown are CE with the pump shapes swapped for $j\in\{0,1\}$. Solid lines are theory (\_th), and markers are experiment (\_ex). Error bars (not shown) are smaller than marker size.}
\label{fig11}
\end{figure}

In Fig. \ref{fig11} we show the CE vs. $\widetilde\gamma$ for our
attempt to match the exact, first input Schmidt mode at every
$\widetilde\gamma$ via numerically computed TM shapes. The plotted
points are for Gaussian, and first-order Hermite-Gaussian pump
pulses. The solid lines are the theoretical prediction. The CE with
matched (appropriate to pump shape) Schmidt mode inputs exceeds those
of the pure Hermite-Gaussian-shaped signal inputs from
Fig. \ref{fig10}. The measured data falls short of theory at larger
pump powers. We suspect this is due to pump pulse reshaping within the
waveguide due to spurious pump-only second-harmonic generation (see
Fig. \ref{fig08}). Despite this, we achieve a CE contrast of $4.7$ to
$1$ ($85$\% vs. $18$\%) between the two pump shapes and their
corresponding, first Schmidt modes. The error bars are smaller than
the plot markers, and are not shown.

\section{Conclusion}

We have quantified and demonstrated the importance of group-velocity
and temporal-width parameter regimes that enhance temporal-mode
selectivity in nonlinear optical frequency conversion in
$\chi^{(2)}$-waveguides. We have confirmed Kumar/Kanter group's
\cite{Manurkar:16} observation of the viability of using standard
frequency-doubling waveguides around the SHG wavelengths, and shown
one such case functioning at near room temperature. And we have
illustrated the use of a single, large, Treacy-grating-pair pulse
shaper, in conjunction with an ultrafast, titanium-sapphire laser and
a reflective spatial-light modulator, to manipulate both the pump and
signal fields, shapes and time delays. And we have demonstrated the
accuracy of the coupled-mode model equations in predicting the
experimental results.

The mode-separability achieved at $50$\% conversion is significant
enough to advance to cascaded frequency conversion schemes
\cite{Reddy:2014bt,Reddy:15pra,Quesada:16}, and study the gains in
high-CE separability from temporal-mode interferometry, thus bringing
us a step closer to a highly-selective quantum pulse gate (QPG)
\cite{EcksteinA:2011vg,Brecht:2011hz,Reddy:2013ip,Brecht:15prx}. The
use of off-the-shelf waveguides and other ubiquitous resources should
empower other groups without access to waveguide fabrication
facilities to invest in temporal-mode selective frequency conversion
research, which is sorely needed in this new, rapidly evolving field.

\section{Acknowledgements}

We would like to thank Benjamin Brecht, and Prof. Christine Silberhorn
for thoughtful discussions and help with system design. We thank Phil
Battle and David Walsh from AdvR for providing the waveguides and
replying to our queries. We acknowledge the support of Larry Scatena
for help with constructing the ultrafast laser, as well as Cliff Dax
and Jeffrey Graman for designing and building the waveguide ovens and
temperature controllers. Lastly, we thank Colin McKinstrie for the
primary theoretical insights that resulted in the conception of this
publication.

Both DVR and MGR received funding from the National Science Foundation
(NSF) grant/award number 1521466, QIS - Quantum Information Science
Program.

\end{document}